\documentclass[]{aastex631}

\let\splitbox\relax

\usepackage{multirow}
\usepackage{graphicx}
\usepackage{adjustbox}
\usepackage{chemformula}
\usepackage{amsmath}
\usepackage{gensymb}
\usepackage{comment}

\begin{document}

\title{Clouds can enhance direct imaging detection of \ch{O2} and \ch{O3} on terrestrial exoplanets}

\author{Huanzhou Yang}
\affiliation{Department of the Geophysical Sciences, University of Chicago}

\author{Michelle Hu}
\affiliation{University of Chicago}

\author{Dorian S. Abbot}
\affiliation{Department of the Geophysical Sciences, University of Chicago}

\begin{abstract}
Clouds are often considered a highly uncertain barrier for detecting biosignatures on exoplanets, especially given intuition gained from transit surveys. However, for direct imaging reflected light observations, clouds could increase the observational signal by increasing reflected light. Here we constrain the impact of clouds on the detection of \ch{O2} and \ch{O3} by a direct imaging telescope such as the Habitable Worlds Observatory (HWO) using observations simulated with the Planetary Spectrum Generator (PSG). We first perform sensitivity tests to show that low clouds enhance \ch{O2} and \ch{O3} detectability while high clouds diminish it, and the effect is greater when cloud particles are smaller.  We next apply clouds produced by the cloud microphysics model CARMA with varied planetary parameters and clouds drawn from observations of different types of clouds on Earth to PSG. 
We find that clouds are likely to increase the SNR of \ch{O2} and \ch{O3} for terrestrial exoplanets under a wide range of scenarios. This work provides important constraints on the impact of clouds on observations by telescopes including HWO.
\end{abstract}


\section{Introduction}

A grand challenge question in exoplanet astronomy is whether life exists on exoplanets. To address this question, NASA is planning the Habitable Worlds Observatory mission (HWO), with a hoped-for launch date in the 2040s \citep{national2021pathways}. HWO will be a direct-imaging telescope that will measure reflected light from the exoplanet itself in the UV, visible, and near-infrared bands. It will have the capability to discover and interrogate Earth-like planets orbiting Sun-like stars, which is not possible for existing missions such as HST, TESS and JWST \citep{sing2011hubble, ricker2015transiting, mather2003james}. As a project like HWO is being planned, it is essential to fully investigate important aspects of its science mission and potential limitations of observing capabilities.

A primary way HWO will be used to assess whether a planet has life is by searching for biosignature gases in its atmosphere. \ch{O2} is a classic biosignature candidate, since it is produced primarily by aerobic respiration on Earth and would not be present in the atmosphere of Earth in significant quantities without life \citep{meadows2018exoplanet,sagan1993search}. The existence of \ch{O3}, another widely-recognized biosignature candidate, requires the presence of \ch{O2} on an Earth-like planet, but it may be easier to detect than \ch{O2} itself due to the location and depth of its absorption bands and because similar levels of \ch{O3} occur even if the concentration of \ch{O2} is 100 times lower than it is on modern Earth \citep{kasting1980evolution}. Like all potential biosignatures, \ch{O2} and \ch{O3} are not perfect, and their detection would not constitute irrefutable evidence that a planet is inhabited \citep{meadows2018exoplanet}. Nevertheless \ch{O2} and \ch{O3} would be strong indicators of the presence of life that are likely to be searched for with HWO, and we will focus on them in this paper.

A critical issue that has affected past observations is the presence of clouds and aerosols in a planet's atmosphere. For example, atmospheric characterization via transmission spectroscopy has been severely limited by clouds \citep{kreidberg2014clouds, sing2016continuum, madhusudhan2019exoplanetary}. Even a thin layer of high cloud can completely disrupt transmission spectroscopic measurements due to the long path length of the transmitted light through the atmosphere. The effect of clouds on reflected light spectroscopy, however, is likely to be more complex. Earth bulk cloud reflectivity data \citep{kawashima2019theoretical} and global climate model simulations \citep[with parameterized clouds and cloud microphysics,][]{checlair2021theoretical} indicate that while a thick cloud deck will block remote access to any target gas located below it, it will also increase the planetary albedo and facilitate the detection of any target gas located above it. This means that cloud height is a critical variable, with high clouds likely to reduce gas detectability, and low clouds likely to increase it \citep{kawashima2019theoretical,checlair2021theoretical}. 

The purpose and novelty of this paper is to investigate in detail the effect of cloud microphysics and vertical distribution on the detectability of \ch{O2} and \ch{O3} by HWO for a terrestrial planet over a range of planetary parameters. We perform our analysis using observations simulated by the Planetary Spectrum Generator (PSG; \citealp{villanueva2022fundamentals}), a radiative transfer model suite designed to synthesize planetary spectra and perform simulated retrievals on them. We apply a variety of different cloud assumptions to PSG: (1) sensitivity tests where we vary cloud height and particle size (microphysics), (2) clouds generated by the cloud microphysics model Community Aerosol and Radiation Model for Atmospheres (CARMA, \citealp{turco1979one, toon1988multidimensional}) for terrestrial planets over a range of planetary parameters \citep{yang2024impact}, and (3) different cloud types observed on Earth.
We find that, for almost all plausible scenarios, clouds tend to improve the \ch{O2} and \ch{O3} detectability for HWO observations. 

This paper is organized as follows. In Section \ref{sec:method}, we describe the models we use and how we conduct simulations of observations. In Section \ref{sec:result}, we describe the mechanism for how clouds impact signal-to-noise ratios (SNR) in observations, and show simulated biosignature detection results with both the simulated and observed clouds. In Section \ref{sec:discussion}, we discuss the limitations of our work and potential future work. We summarize our results in Section \ref{sec:summary}.

\section{Model Description}\label{sec:method}

\subsection{Planetary Spectrum Generator}

We use PSG to simulate observations with a 6-meter segmented on-axis telescope as outlined in \citet{national2021pathways}. We use built-in PSG detailed settings for simulated observations with LUVOIR-A, the closest available option to HWO (Table \ref{tab:psgset}). Compared to transit targets, one advantage of direct imaging is that it allows for more possible alignments of the planetary orbit. 
The phase angle is defined as the star-planet-observer angle, which is used to describe the orientation of the targeted stellar system. The planet is at primary eclipse when the phase angle is $180\degree$ and at secondary eclipse when the phase angle is $0\degree$. The maximum and minimum observable phase angles are determined by the inner working angle (IWA) of the star shade, which blocks light from the star to allow direct imaging of the planet \citep{bolcar2018large}. \cite{national2021pathways} recommended that the IWA be less than 60 mas. For a stellar system at a distance of 5 pc, this corresponds to a projected star-planet distance of about 0.3 AU, which means the minimum phase angle for a successful observation of a planet orbiting at 1~AU is about $17\degree$. For the main set of experiments, we use a phase angle of $90\degree$, for which the exozodiacal noise is minimized. Our main conclusions in Section \ref{sec:result} concerning the effect of clouds are robust to variations in the phase angle.

We evaluate the detectability of the biosignatures \ch{O2} and \ch{O3} using the signal-to-noise ratio. The signal-to-noise ratio as a function of wavelength (SNR$_\lambda$) is defined as
\begin{equation}
    \text{SNR}_\lambda = \frac{\text{Signal}}{\text{Noise}}= \frac{I_{\lambda, without}-I_{\lambda, all}}{I_{\lambda, noise}},
    \label{snr_define1}
\end{equation}
where $I_{\lambda, all}$ is the spectral intensity of radiation from a planet, $I_{\lambda, without}$ is the spectral intensity of radiation from a planet without the targeted biosignature gas, and $I_{\lambda, noise}$ is the spectral intensity of radiation caused by the following sources of noise: exozodi, instrumental optics, and other background sources \citep{villanueva2022fundamentals}. We then calculate the overall signal to noise ratio (SNR) as 
\begin{equation}
    \text{SNR} = \sqrt{\sum_\lambda (\text{SNR}_\lambda)^2}.
    \label{snr_define2}
\end{equation}
We calculate SNR in clear-sky and cloudy scenarios, and then compare them to determine the impact of clouds. A conventional criterion for gas detectability is an SNR of 5, while an SNR of 3 can be used as a threshold for preliminary detection \citep{lustig2019detectability,meadows2023feasibility}. Here we use an integration time of 3 hours. The SNR results from PSG can be rescaled by the square root of the integration time, and the sign of the effect of clouds on the SNR do not depend on the integration time assumed.

\begin{table}[htbp]\label{tab:psgset}
    \centering
    \begin{adjustbox}{max width=\textwidth}
    \begin{tabular}{c|c}
        \hline
        \textbf{Simulation Setup} & \textbf{Value} \\
        \hline
        Phase angle & 90$^{\circ}$ \\ 
        \hline
        Target star distance  & 5 pc \\ 
        \hline        
        Integration time  & 3 hour \\ 
        \hline        
        Wavelengths  & 0.2-2 $\mu m$ (UV, VIS, near NIR)\\ 
        \hline 
        Resolving power & UV: 7; VIS: 140; NIR: 70\\ 
        \hline 
        Exozodi level & 4.5 \\
        \hline
        Surface albedo & 0.15 \\
        \hline
    \end{tabular}
    \end{adjustbox}
    \caption{Description of the experimental setup in PSG, which is the model we use for the simulated observations. A phase angle of 90$^{\circ}$ is halfway between primary and secondary eclipse. The exozodi level is relative to the Solar System.}
\end{table}

\subsection{Cloud Data from CARMA}

\begin{figure*}
    \centering
    \includegraphics[width=0.8\textwidth]{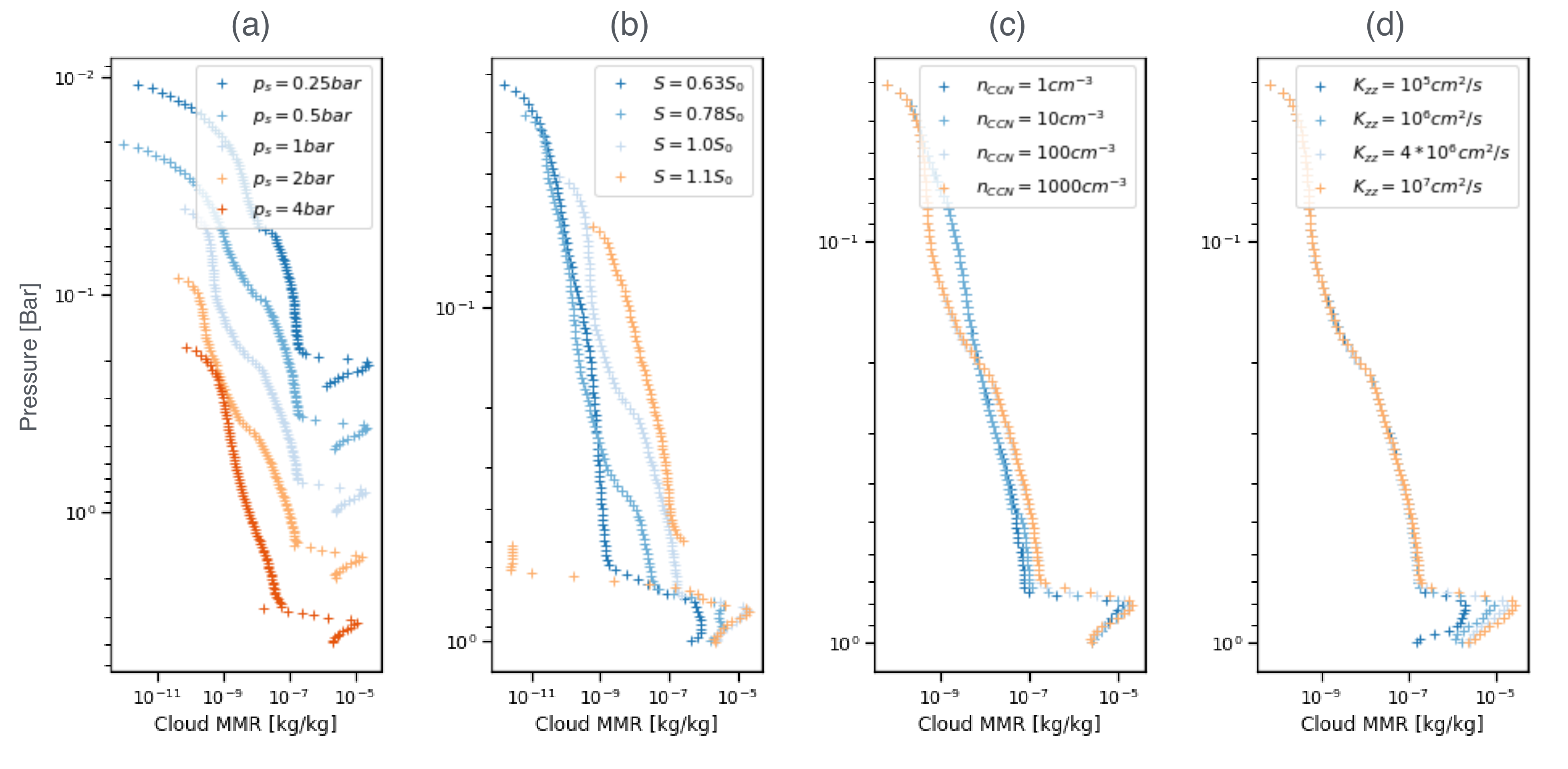}
    \caption{Vertical profiles of cloud mass mixing ratio (MMR) with different planetary parameters. The four parameters are: surface pressure (a), stellar irradiation (b), number density of cloud condensation nucleus (c, CCN) and vertical diffusivity in the mixed layer (d, $K_{zz}$). Data are from \citet{yang2024impact}.}
    \label{fig:mmr}
\end{figure*}

\begin{figure*}
    \centering
    \includegraphics[width=0.8\textwidth]{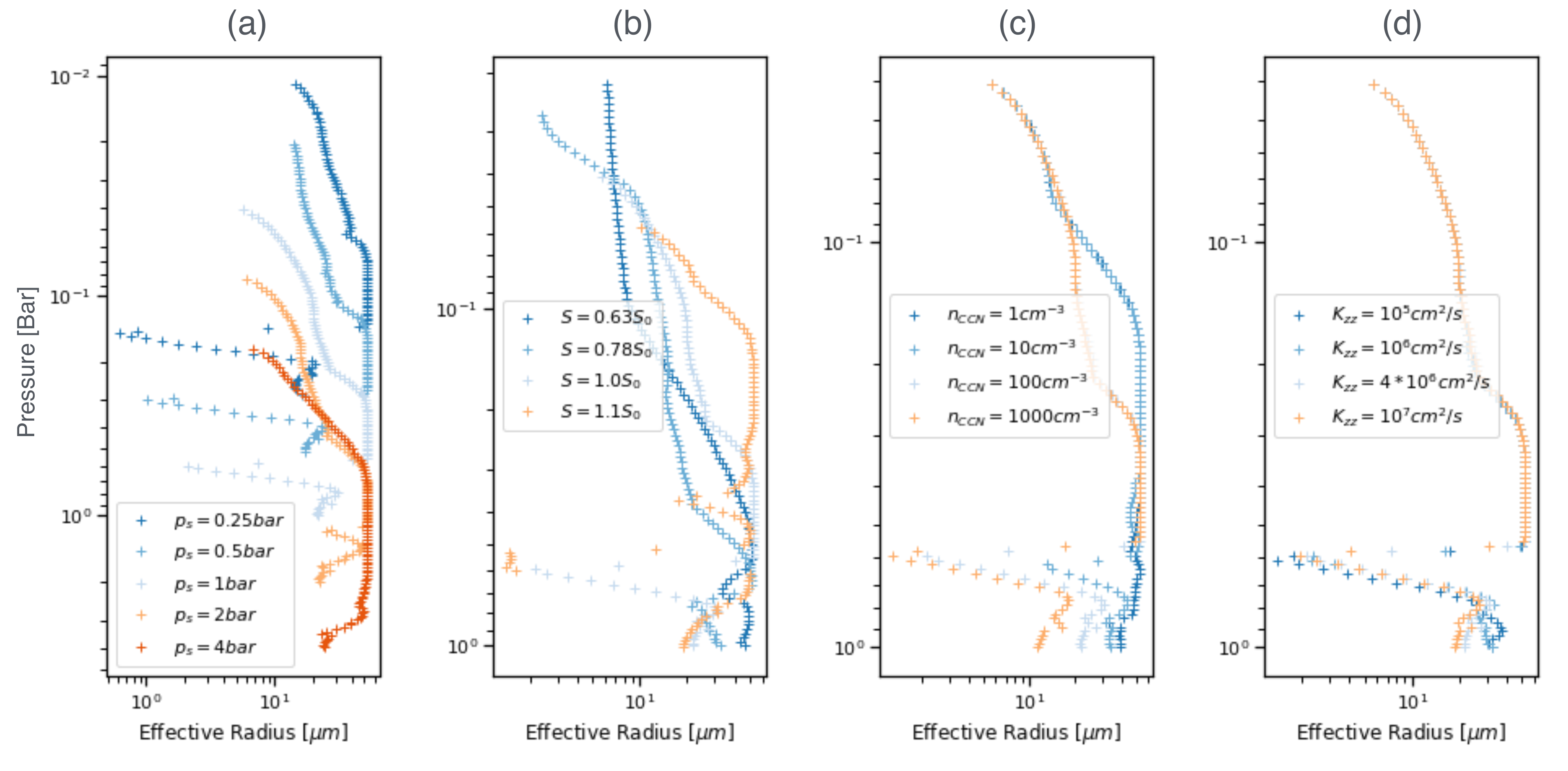}
    \caption{Vertical profiles of cloud effective radius ($R_{eff}$) with varying planetary parameters. The four parameters are: surface pressure (a), stellar irradiation (b), number density of cloud condensation nucleus (c, CCN) and vertical diffusivity in the mixed layer (d, $K_{zz}$).  Data are from \citet{yang2024impact}.}
    \label{fig:reff}
\end{figure*}

The Community Aerosol and Radiation Model for Atmospheres (CARMA) is a cloud microphysics model that explicitly simulates the vertical and size distribution of cloud particles. \cite{yang2024impact} conducted 1-D CARMA simulations over a wide range of planetary parameters. The resulting clouds were most similar to stratocumulus and stratus clouds on Earth \citep{yang2024impact}. Of the 12 planetary parameters they tested, the following parameters had the largest impact on cloud properties: surface pressure, stellar irradiation, number density of cloud condensation nuclei (CCN) and vertical diffusivity in the mixed layer ($K_{zz}$). Planetary surface pressure determines the pressure of the main cloud deck because there is always a low cloud deck near surface (Fig.~\ref{fig:mmr}a).  Higher surface pressure also increases the cloud particle radius (Fig.~\ref{fig:reff}a) by decreasing their terminal velocity. Stellar irradiation affects the temperature profile of the atmosphere and thus the vertical and size distributions of clouds (Fig.~\ref{fig:mmr}b \& \ref{fig:reff}b), in a complicated way \citep[see][for a more detailed discussion]{yang2024impact}. Increasing the number density of CCN strongly decreases the cloud radius (Fig.~\ref{fig:reff}c), but does not strongly affect the cloud mass mixing ratio (Fig.~\ref{fig:mmr}c). This is because a similar quantity of cloud is spread over more cloud droplets, each nucleated around a CCN.  Increasing the vertical diffusivity in the boundary layer increases the supply of condensable water vapor from the surface to the cloud layer, and therefore significantly increases the mass mixing ratio of clouds (Fig.~\ref{fig:mmr}d), while slightly decreasing the cloud radius because increased vertical diffusivity supplies more CCN to the cloud layer (Fig.~\ref{fig:reff}d).

We process the CARMA output with PSG to simulate observations. The PSG default is to use a two-moment description of mass mixing ratio and effective radius for atmospheric aerosols, including water clouds. Various built-in Mie-scattering datasets can be applied to these clouds \citep{villanueva2022fundamentals}. As we need to resolve cloud size distributions in a bin model, we created a user-defined cloud species from CARMA output data. With the refractive indices of water and ice, CARMA provides an output of Mie-scattering properties for clouds of different sizes. PSG has a limit of 20 customizable aerosol types, but we are running CARMA with 100 vertical levels. As a result, when we input CARMA output into PSG, we treat clouds over multiple adjacent levels representing $\sim$1~km of altitude as one species, and calculate the combined Mie-scattering properties as 
\begin{align}
    q_{ext} &= \frac{\sum_{r}\pi r^2n(r)Q_{ext}(r)}{\sum_{r} \pi r^3n(r)\rho_{cloud}} \label{equ:3}\\
    q_{sca} &= \frac{\sum_{r}\pi r^2n(r)Q_{ext}(r)\alpha_{ssa}(r)}{\sum_{r} \pi r^3n(r)\rho_{cloud}} \label{equ:4}\\
    g_{asym} &= \frac{\sum_{r}\pi r^2n(r)Q_{ext}(r)\alpha_{ssa}(r)g(r)}{\sum_r \pi r^2n(r)Q_{ext}(r)\alpha_{ssa}(r)},
\end{align}
where $q_{ext}$, $q_{sca}$ and $g_{asym}$ are the mass extinction coefficient, mass scattering coefficient and asymmetry factor of the combination of cloud particles with different sizes, $r$ denotes the radius of cloud particles, $n(r)$ denotes the number density of cloud particles as a function of radius, and $Q_{ext}$, $\alpha_{ssa}$ and $g$ are extinction coefficient, single scattering albedo and asymmetry factor of clouds of a single size directly output from CARMA. The numerator in Equation \ref{equ:3} gives the total cloud cross section \citep[$\sigma_{sca} = \pi r^2n(r)Q_{ext}(r)\alpha_{ssa}(r)$,][]{lamb2011physics}, and the denominator gives the mass of the cloud. The mass scattering coefficient (Equation \ref{equ:4}) is defined similarly. The asymmetry parameter can be obtained by the average of the asymmetry parameter of each size bin weighted by the scattering cross section \citep{wolf2004mie}.

When we vary the surface pressure, $P_s$, in PSG we do it by changing the \ch{N2} partial pressure. When the surface pressure varies, we adjust the mass mixing ratio of \ch{O2} to keep a constant column mass. We do this by maintaining the \ch{O3} partial pressure constant above 0.1 bar (\ch{O3} is negligible below 0.1 bar). This assumption that the ozone layer forms at a constant pressure is reasonable given that the position of ozone is mainly determined by its production from solar UV and its destruction through catalytic cycles, which are primarily controlled by the UV optical depth above the ozone layer, which is roughly proportional to pressure \citep{seinfeld2016atmospheric}. When we vary the surface pressure, we apply water vapor concentrations to PSG taken from CARMA experiments. The column mass of water vapor does not vary strongly with surface pressure because it is limited by vertical transport.
When we process clouds from CARMA experiments with different stellar irradiation, we leave the stellar irradiation constant when we process the CARMA output with PSG. The reason is that higher stellar radiation trivially leads to larger SNR, and we want to focus our analysis on the effect of changes in cloud properties. 

\section{Results}\label{sec:result}

\subsection{Factors that influence cloud effect on SNR}\label{sec:sensi}

\begin{figure*}
    \centering
    \includegraphics[width=0.8\textwidth]{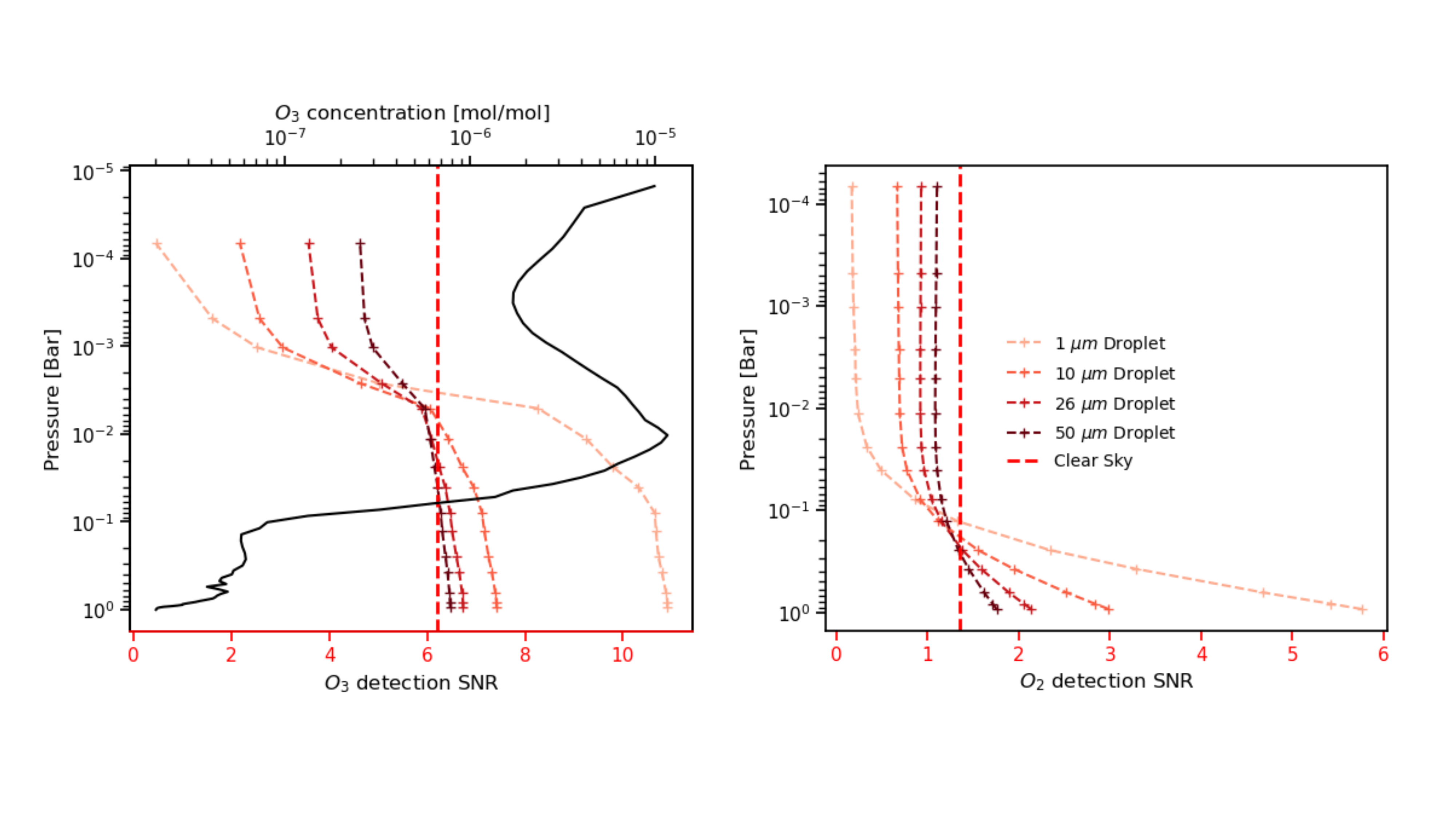}
    \caption{The impact of cloud height and particle size on SNR. The left panel is for $\text{O}_3$ detection, and the right panel is for $\text{O}_2$. We tested cloud particles with radii of 1 $\mu m$, $10 \mu m$, $26 \mu m$ and $50 \mu m$. The vertical dashed line shows the SNR in the clear-sky case. We also plot the vertical concentration profiles of $\text{O}_3$ in solid black for reference.}
    \label{fig:sensi}
\end{figure*}

First we examine the impact of cloud height and particle size in a group of control experiments.
In each experiment, we add monodisperse (single-sized) cloud to a single pressure level of the atmosphere.
These experiments show us the sensitivity of biosignature gas detectability to cloud height and particle size. The results are plotted in Figure \ref{fig:sensi}, where each data point represents the SNR from one sensitivity experiment.
We only show results for liquid droplets here because we assume ice particles are spherical in this work and therefore have similar optical features as liquid particles.
We test four cloud particle radii: 1 $\mu m$, 10 $\mu m$, 26 $\mu m$ and 50 $\mu m$. We can draw the following conclusions from the figure:
\begin{enumerate}
    \item\label{itemone} Lower clouds are more likely to increase the detectability of biosignatures and higher clouds are more likely to decrease the detectability.
    \item\label{itemtwo} With the same cloud water content, smaller cloud particle sizes magnify the impact of the cloud.
   \item\label{itemthree} There is a `neutral height' where the existence of clouds does not impact SNR. This `neutral height' gets lower with larger cloud particle size.
\end{enumerate}

Qualitatively, these conclusions hold for both \ch{O2} and \ch{O3}, although quantitatively the percentage change in SNR and the altitude of the `neutral height' differ between the gases. Conclusions \ref{itemone} and \ref{itemtwo} match the physical intuition described in \cite{kawashima2019theoretical,checlair2021theoretical}. Specifically: (1)
clouds can block gas absorption features below the cloud level, while they can magnify the information from gas absorption above the cloud level due to their high reflectivity and (2) larger cloud albedo caused by smaller cloud radius increases the impact of clouds. 
In order to understand all three points better, we develop a simple analytical model to describe the impact of clouds on SNR.

\begin{figure*}
    \centering
    \includegraphics[width=0.6\textwidth]{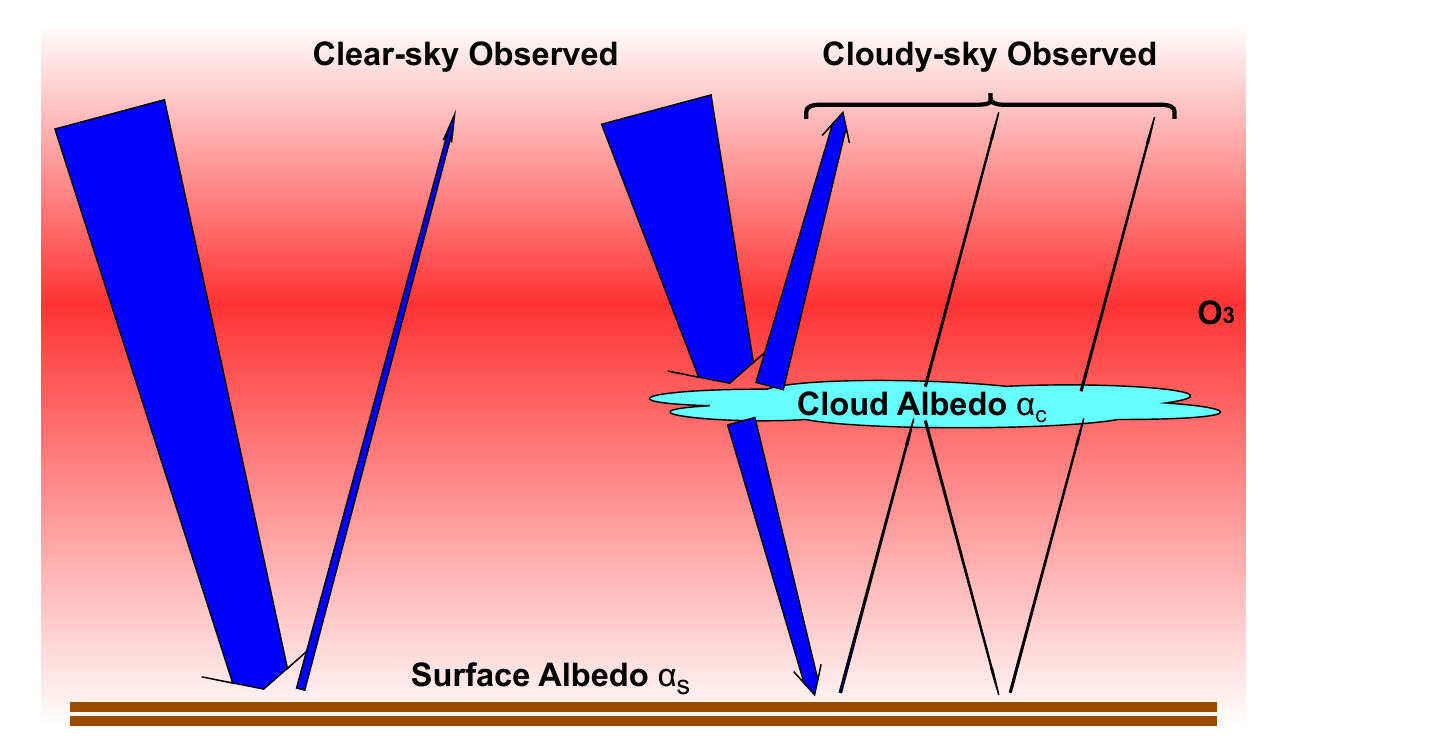}
    \caption{Schematic diagram of the analytical model (Eq.~(\ref{equ:cloud})) we use to understand the effect of cloud height and particle size on SNR. The left part of the diagram represents clear-sky and the right represents cloudy-sky. The solid horizontal line at the bottom of the diagram represents the planet's surface, which has albedo (reflectivity) $\alpha_s$. The cloud layer has albedo $\alpha_c$ and zero absorptivity. The arrows represent light moving through the atmosphere, with thickness corresponding qualitatively to intensity. The arrows get thinner from tail to head, representing absorption in atmosphere. The background red color shows the assumed vertical distribution of ozone, which is concentrated at a high altitude.}
    \label{fig:analy}
\end{figure*}

Figure \ref{fig:analy} shows a sketch of our simplified analytical model. The planet has a surface with albedo $\alpha_s$ and a cloud layer at height $z_c$ with albedo $\alpha_c$. 
There are two radiatively active gases in the 1-D atmosphere column, the background gas with column-integrated optical depth $\tau_0$ and the target gas with optical depth $\tau_t$. 
We neglect emission from the atmosphere and the planet's surface because the telescope we simulate does not observe wavelengths longer than 2 $\mu m$. Because the single scattering albedo ($\alpha_{ssa}$, the ratio of the scattered light and the light scattered and absorbed) is close to 1 for clouds in the observed waveband, we also neglect cloud absorption, which means that light is either transmitted or reflected by the cloud.
In the analytical model, we apply the gray atmosphere assumption that optical properties are independent of wavelength. Conclusions from the model can be generalized to the full spectrum with Equation~\ref{snr_define2} through the assumption that noise in different wavelength bands is uncorrelated, following PSG. 
We also assume that noise is independent of the light from the target planet, which is valid provided that the planet is faint (as is the case for terrestrials). This allows us to focus on the `signal' component of SNR in our analytical model. The signal is proportional to the light traveling towards the observer. Here we use the upwelling radiation at the top of the atmosphere to represent the signal, neglecting the effect of phase angle and the spherical shape of exoplanets in 3-D, which do not affect our results qualitatively.

In the cloudy-sky case, the upward reflected light at the top of the atmosphere has two components: light directly reflected from the cloud layer and light transmitted through the cloud and reflected by the surface (right part of Figure \ref{fig:analy}). In the clear-sky case, the light we observe is just the reflection from the planet surface (left part of Figure \ref{fig:analy}). The impact of the cloud on detection is determined by the difference between these two values. Here we use $s_{clear}$ and $s_{cloud}$ to denote the signal for detecting the target gas for clear-sky and cloudy-sky cases.

Applying Equation \ref{snr_define1}, the clear-sky signal, $s_{clear}$, normalized by the top-of-atmosphere incoming stellar flux, $I_0$, can be written as:
\begin{align}\label{equ:6}
    s_{clear} &= \frac{I_{without}-I_{all}}{I_0} \\
    &= \frac{I_0\alpha_s e^{-2\tau_0}-I_0\alpha_s e^{-2\tau_0-2\tau_t}}{I_0}\\
    &= \alpha_s e^{-2\tau_0}(1- e^{-2\tau_t})
\end{align}
where $I_{without}$ is the upward radiation at the top of atmosphere with background gas only and $I_{all}$ is the upward radiation at the top of atmosphere with both gases. With cloud present, the upward top-of-atmosphere radiative flux without the target gas is given by:
\begin{equation}
    I_{without} = I_0\alpha_c e^{-2(\tau_0-\tau_{0,b})} + I_0 \alpha_s (1 - \alpha_c)^2 e^{-2\tau_0} \sum_i (\alpha_s \alpha_c e^{-2\tau_{0,b}})^i, 
\end{equation}
where $\tau_{t,b}$ denotes the optical depth of target gas below the cloud level and  $\tau_{0,b}$ denotes the optical depth of background gas below the cloud level, both of which are a function of the cloud height.
The first term is the radiation directly reflected by cloud and the second term is the radiation transmitted first and ultimately reflected upward after reflections between the planetary surface and cloud (Figure \ref{fig:analy}). The radiative flux assuming the target gas is present, $I_{all}$, is calculated in a similar way.
The cloudy-sky signal is therefore:
\begin{align}
     s_{cloud} &= \alpha_c e^{-2(\tau_0-\tau_{0,b})}(1 - e^{-2(\tau_t-\tau_{t,b})}) +  (1 - \alpha_c)^2 \alpha_s e^{-2\tau_0} \left( \sum_i (\alpha_s \alpha_c e^{-2\tau_{0,b}})^i - e^{-2\tau_t} \sum_i (\alpha_s \alpha_c e^{-2(\tau_{0,b}+\tau_{0,b})})^i  \right), \\
    s_{cloud} &= \alpha_c e^{-2(\tau_0-\tau_{0,b})}(1 - e^{-2(\tau_t-\tau_{t,b})}) + (1-\alpha_c)^2\alpha_s e^{-2\tau_0}  (\frac{1}{1-\alpha_s \alpha_c e^{-2\tau_{0,b}}}-\frac{e^{-2\tau_t}}{1-\alpha_s \alpha_c e^{-2(\tau_{0,b}+\tau_{t,b})}}). \label{equ:cloud}
\end{align}
As visualized in Figure \ref{fig:analy}, the first term in Equation \ref{equ:cloud}, representing direct reflection, is the dominant term in most cases due to the high cloud albedo and low surface albedo. This term describes the two main impacts of clouds: enhancing information from the atmosphere above them and blocking information from the lower atmosphere.

\begin{figure*}
    \centering
    \includegraphics[width=0.8\textwidth]{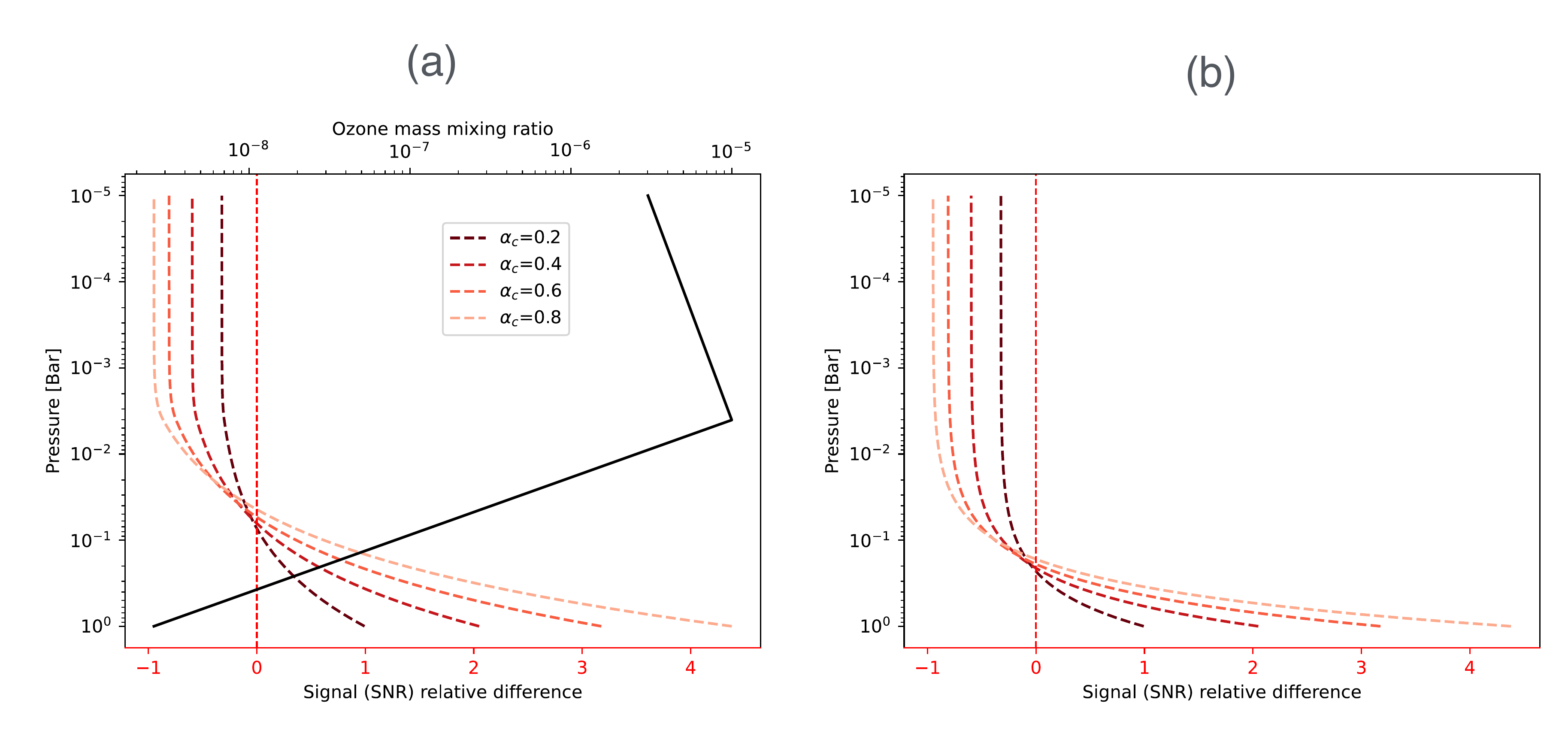}
    \caption{Results from the analytical model for detections of \ch{O3} (a) or \ch{O2} (b). The analytical model results should be compared with Figure \ref{fig:sensi}. The concentration of \ch{O3} is prescribed as a simple vertical distribution function (black curve in (a)). \ch{O2} is assumed to be well-mixed. Similar to Figure \ref{fig:sensi}, the `neutral cloud impact height' decreases with decreased cloud albedo, which corresponds to increased cloud particle size. In Figure \ref{fig:sigtau}a where \ch{O3} is the target gas, $\tau_0 = 0.01$ and $\tau_t=0.1$. In Figure \ref{fig:sigtau}b where \ch{O2} is the target gas, $\tau_0 = 0.1$ and $\tau_t=0.01$.}
    \label{fig:sigtau}
\end{figure*}

Figure \ref{fig:sigtau} shows the results of our analytical model (Eq.~\ref{equ:cloud}), which should be compared with the PSG calculations shown in Figure \ref{fig:sensi}. A larger cloud albedo in our analytical model corresponds to a smaller cloud particle size in Figure \ref{fig:sensi}. In the calculation, we assume that the total optical depth for \ch{O2} is 0.01 and that for \ch{O3} is 0.1. With \ch{O3} as the target biosignature gas in Figure \ref{fig:sigtau}a, $\tau_0 = 0.01$ and $\tau_t=0.1$. In Figure \ref{fig:sigtau}b where \ch{O2} is the target biosignature gas, $\tau_0 = 0.1$ and $\tau_t=0.01$. \ch{O2} has a constant mixing ratio throughout the atmosphere, and the simplified ozone distribution that we assume is shown in Figure \ref{fig:sigtau}a.
This figure reproduces the three conclusions we drew from Figure \ref{fig:sensi}.


To better understand the analytical model results plotted in Figure \ref{fig:sigtau}, we can simplify Equation \ref{equ:cloud}. 
$\alpha_s=0.15$ and $\alpha_c \lesssim 0.7$ for most clouds, therefore $\alpha_s \alpha_c e^{-2\tau_{0,b}}\leq \alpha_s\alpha_c\lesssim0.1$. This allows us to approximate the denominator $1-\alpha_s \alpha_c e^{-2\tau_{0,b}}$ as 1 even with low optical depths for both gases. We then compare $s_{cloud}$ to $s_{clear}$ for the impact of clouds:
\begin{equation}\label{equ:deltas}
    \Delta s_{cloud} = s_{cloud}-s_{clear} = \alpha_c e^{-2(\tau_0-\tau_{0,b})}(1-e^{-2(\tau_t-\tau_{t,b})}) - \alpha_c(2-\alpha_c)\alpha_s e^{-2\tau_0}  (1-e^{-2\tau_t}).
\end{equation}
First let's conduct an asymptotic analysis of this equation. 
In the limit without clouds ($\alpha_c \rightarrow 0$), we have a trivial solution where  $\Delta s_{cloud} = 0$. In the limit of extremely optically thick clouds ($\alpha_c \rightarrow 1$), the first term is equivalent to a perfectly reflecting surface at cloud height, and the second term reduces to the clear-sky signal.
In the limit of clouds at the surface ($z_c \rightarrow 0$), we have:
\begin{equation}\label{equ:low}
    \Delta s_{cloud} = (1-(2-\alpha_c)\alpha_s) \alpha_c e^{-2\tau_0}  (1-e^{-2\tau_t})
\end{equation}
In this limit clouds will benefit detections as long as $\alpha_c>2-\frac{1}{\alpha_s}$. This means that any amount of cloud will increase detectability for $\alpha_s<0.5$, corresponding to any planet that is not mostly covered by ice and snow. For an Earth-like planet with a surface albedo of 0.15, we can be confident that near-surface clouds would enhance biosignature detection. 
Alternatively, if clouds approach the top of the atmosphere ($\tau_{t,b} \rightarrow \tau_t$ and $\tau_{0,b} \rightarrow \tau_0$), the first term becomes zero, and we have:
\begin{equation}\label{equ:high}
    \Delta s_{cloud} = - \alpha_c(2-\alpha_c)\alpha_s e^{-2\tau_0}(1-e^{-2\tau_t}) < 0
\end{equation}
This means the highest clouds will always reduce SNR. These high and low cloud limits reproduce conclusion \ref{itemone} at the beginning of this section. In Equation \ref{equ:low}, $\Delta s_{cloud}$ increases with $\alpha_c$. In Equation \ref{equ:high}, the magnitude of $\Delta s_{cloud}$, which is negative, increases with $\alpha_c$. Since a higher cloud albedo corresponds to a smaller cloud radius, this explains conclusion \ref{itemtwo}. 

To show conclusion \ref{itemthree} with the analytical model, we derive the neutral cloud impact height, $z_c^*$, by setting $\Delta s_{cloud}=0$ in Equation \ref{equ:deltas}, where $\tau_{t,b}$ and $\tau_{0,b}$ are functions of cloud height, $z_c$. This leads to the following equation for $\alpha_c$
\begin{align}
    \alpha_c&= 2- \frac{1}{\alpha_s(1-e^{-2\tau_t})}\frac{1-e^{-(2\tau_t-2\bar k_{t,b}z_c^*)}}{e^{-2\bar k_{0,b}z_c^*}} \label{equ:zstar}
\end{align}
where $\tau_{t,b}=\bar k_{t,b}z_c$ and $\tau_{0,b}=\bar k_{0,b}z_c$ are represented with the height of cloud and the average extinction $\bar k_{t,b}$ and $\bar k_{0,b}$ below cloud level.
The optically thin limit ($\tau_0 \ll 1$ and $\tau_t\ll1$) is a reasonable approximation for terrestrial planets. Making this assumption, we can approximate $e^{-\tau}$ as $1-\tau$ for all such terms, which leads to
\begin{align}
    z_c^* &= \frac{1}{2\bar k_0} (1+\frac{\frac{\bar k_t}{\bar k_0}-2\tau_t}{\alpha_s(1-e^{-2\tau_t}) (2-\alpha_c) - \frac{\bar k_t}{\bar k_0}})
\end{align}
For gases with roughly similar vertical distributions, $\frac{\bar k_t}{\bar k_0}\sim\frac{\tau_t}{\tau_0}$, such that $\frac{\bar k_t}{\bar k_0}\gg\tau_t$ when $\tau_0\ll1$. According to our assumptions, therefore, $\frac{\bar k_t}{\bar k_0}-\tau_t>0$. We then have that $z_c^*$ increases with $\alpha_c$, corresponding to decreases in cloud particle size, which explains conclusion \ref{itemthree}. It is important to note, however, that conclusion \ref{itemthree} is less strong than conclusions \ref{itemone} and \ref{itemtwo}, in that this derivation depended on assumptions about the vertical distribution of absorbers and the optically thin approximation.

\subsection{Effect of planetary parameters}
\label{sec:planetary_params}

In Section \ref{sec:sensi}, we performed sensitivity tests to investigate how cloud particle radius and height impact SNR. In this section, we feed resolved cloud radius and height output from 1D CARMA with different planetary parameters \citep{yang2024impact} into PSG to assess the cloud effect on SNR for specific planetary scenarios. 

\begin{figure*}
    \centering
    \includegraphics[width=0.8\textwidth]{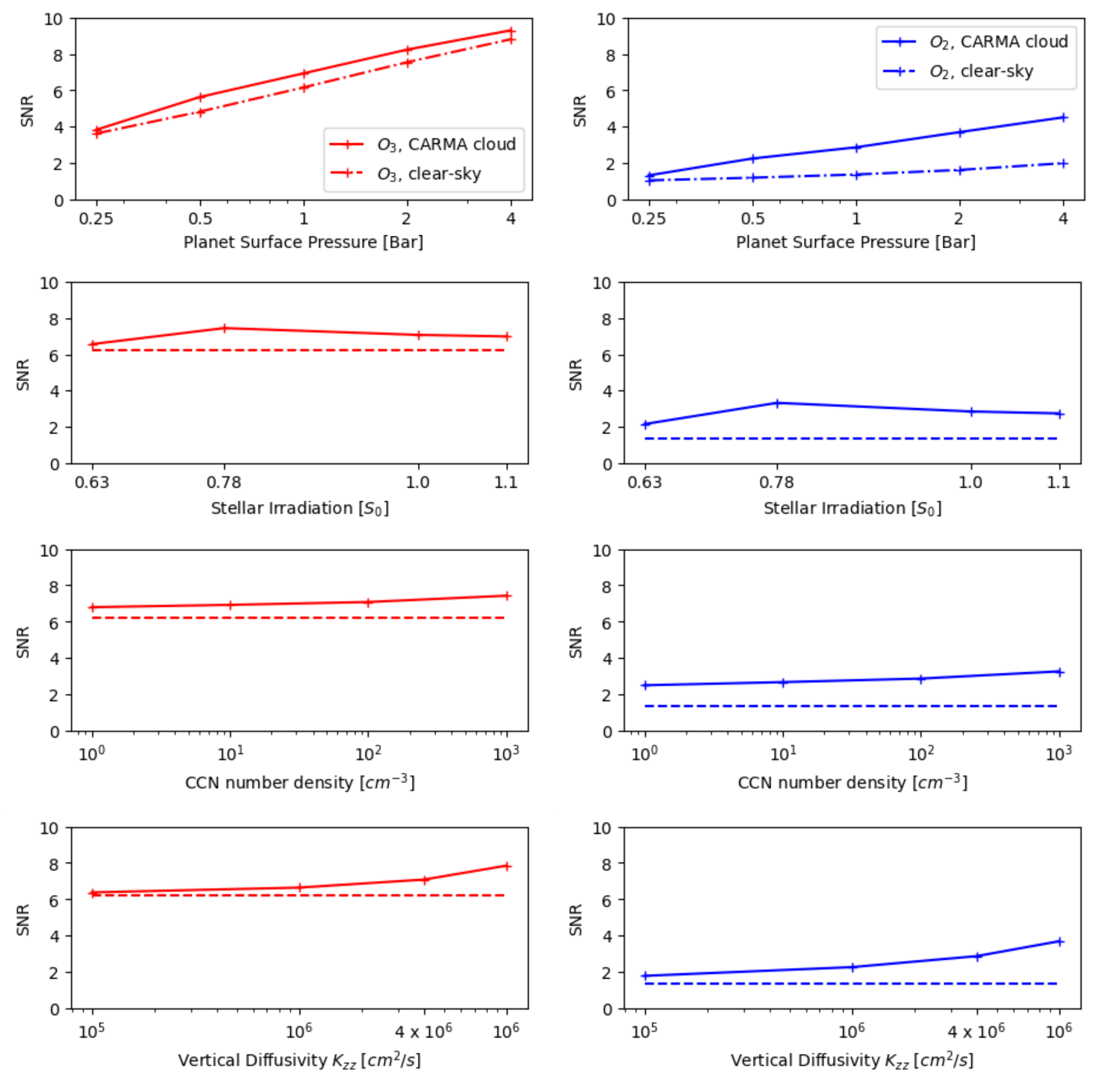}
    \caption{Cloud impact on gas detection when planetary parameters are varied. The cloud properties and vertical distribution for a given set of planetary parameters are calculated using CARMA \citep{yang2024impact}. The left column is for \ch{O3} detection, and the right column is for \ch{O2}. From top to bottom, the rows show the result of varying planetary surface pressure, stellar irradiation, CCN number density and vertical diffusivity. Solid lines are for simulations with clouds while dashed lines are for clear-sky simulations. The clear-sky baseline simulation varies with surface pressure, but not the other parameters.}
    \label{fig:param}
\end{figure*}

Figure \ref{fig:param} summarizes how the cloud scenarios simulated by CARMA affect the SNR for detecting \ch{O2} and \ch{O3} simulated by PSG. There are a few qualitative observations we can make from these results that apply to all planetary scenarios. First, including clouds increases the SNR in all cases. This is because clouds increase the planetary albedo, and therefore the number of photons for detection. Second, \ch{O3} is generally a better biosignature with higher SNR, owing to its strong absorption band in the UV. Third, the fractional effect of clouds on SNR is larger for \ch{O2} than \ch{O3}. The reason is that the main \ch{O3} absorption bands are in the UV, and the clear-sky planetary albedo is already high in the UV due to strong Rayleigh scattering, so adding clouds increases the albedo less.

The specific planetary parameters have some unique effects on the cloud impact on SNR. Increasing surface pressure increases the SNR of \ch{O3} and \ch{O2} both with or without clouds (Fig.~\ref{fig:param}, top row). Since we fixed the column mass of these gases, this increase in detectability is not due simply to an increase in gas abundance. Instead, the SNR increases because Rayleigh scattering increases with higher surface pressure, resulting in increased planetary albedo. Note that the clear-sky SNR increases much more with surface pressure for \ch{O3} than for \ch{O2}. This is because, as noted above, \ch{O3} absorbs much more strongly at the short wavelengths that are most affected by Rayleigh scattering. For \ch{O2}, clear-sky SNR is relatively constant as surface pressure is varied because its absorption band is not concentrated at short wavelengths where the albedo is strongly affected by Rayleigh scattering. \ch{O2} SNR increases more significantly as surface pressure increases when clouds are included. This is because cloud water content (and therefore albedo) increases strongly with surface pressure due to more condensible water vapor both from increased surface temperature and increased diffusion from the surface \citep{yang2024impact}. 

When varying stellar irradiation, we find that clouds have a small impact on SNR at low irradiation (Fig.~\ref{fig:param}, second row). This is because there are only ice clouds at low temperatures, which have a reduced mass mixing ratio (Fig.~\ref{fig:mmr}b) because the formation of ice particles is less efficient than liquid droplets in the model \citep{yang2024impact}. Additionally, clouds have a reduced impact on SNR at high stellar irradiation. This is because ice clouds move to higher altitudes when the surface temperature is higher (Fig.~\ref{fig:mmr}b), where they block more of the atmosphere to remote observation. Increasing CCN number concentration reduces the cloud effective radius (Figure \ref{fig:reff}), and therefore increases the cloud albedo and the cloud impact on SNR (Fig.~\ref{fig:param}, third row). Increasing vertical diffusivity $K_{zz}$ increases the cloud amount (Fig.~\ref{fig:reff}d) and therefore increases the cloud albedo and cloud impact on SNR (Fig.~\ref{fig:param}, bottom row).


\subsection{Effect of cloud types}

Clouds on Earth are categorized into different types such as stratus, cumulus, and cirrus. Each type of cloud has a typical distribution of height, phase (liquid, ice, or mixed), and particle size. Cloud formation is determined by a variety of atmospheric processes, many of which cannot be resolved by 1D CARMA simulations. The clouds that 1D CARMA forms tend to be most similar to stratus and stratocumulus \citep{yang2024impact}.

To investigate the effect of different cloud types on biosignature retrievals, we apply tropical ice and liquid cloud content from observational data collected by \citet{huang2015climatology} to PSG. \citet{huang2015climatology} estimate the ice cloud content uncertainty is around 30\% \citep{deng2013evaluation} and liquid clouds at low-level or with small particle sizes can be missed due to the lack of lidar-measurements \citep{sassen2008classifying}. These uncertainties should not affect the qualitative comparisons of the effect of different cloud types we make here. We consider the five types of clouds that are most distinct in their vertical distribution: St (Stratocumulus + Stratus), Ns (Nimbostratus), Dc (Deep Convective), Cu (Cumulus) and Ci (Cirrus). Each type of cloud can be composed of ice clouds, liquid clouds, or both. Cloud sizes are not explicitly given in \citet{huang2015climatology}. Measurements from different sources show a range of effective radii for clouds, depending on time, location, and method of measurement. Effective radii of liquid cloud droplets are typically $5-15~\mu m$ \citep{martin1994measurement, miles2000cloud, brenguier2000radiative}. Ice crystal effective radii are $10-50~\mu m$ \citep{baum2005bulk, seinfeld2016atmospheric, krisna2018comparing}. We test this full range by conducting experiments both for large cloud particle radii ($15~\mu m$ for liquid droplets and $50~\mu m$ for ice crystals) and small cloud particle radii ($5~\mu m$ for liquid droplets and $10~\mu m$ for ice crystals). 

\begin{figure*}
    \centering
    \includegraphics[width=0.8\textwidth]{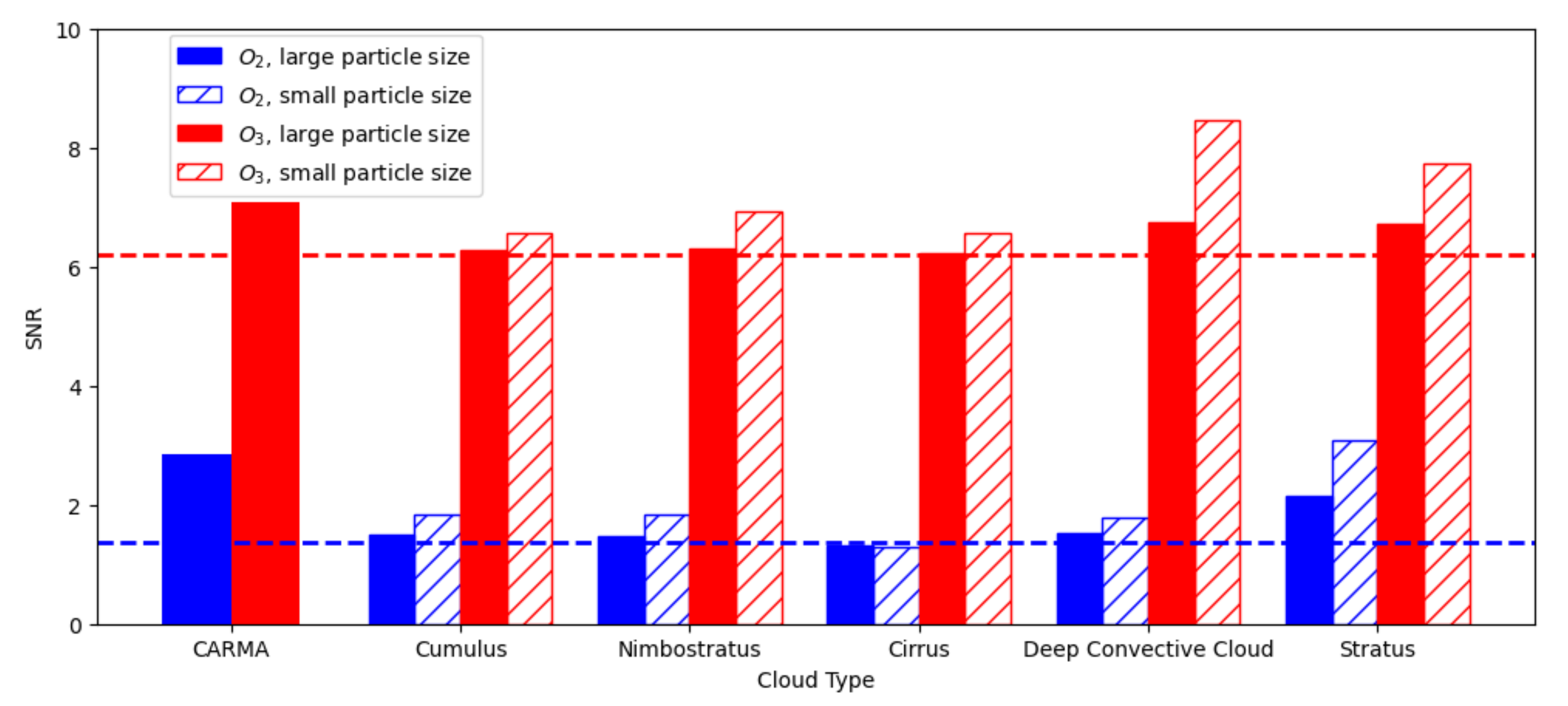}
    \caption{SNR for detection of \ch{O2} (blue) and \ch{O3} (red) with different types of clouds present in the atmosphere of an Earth-like planet. Dashed lines represent clear-sky results and bars represent cloudy results (solid assuming large cloud particles and hatched assuming small cloud particles). CARMA models cloud particle size, so cloud sizes do not need to be assumed.}
    \label{fig:cloudtype}
\end{figure*}

In Figure \ref{fig:cloudtype}, we show the SNR for detecting \ch{O2} and \ch{O3} assuming different cloud types. The most important point is that for all cloud types and both extremes of the particle size distribution, clouds have either a neutral or positive effect on detectability. High-altitude cirrus clouds slightly decrease the SNR for \ch{O2} detection, but this effect is very small because of their small water content.  Since clouds have a neutral or positive effect on SNR, and decreasing the cloud particle size increases the cloud effect on SNR (Fig.~\ref{fig:sensi}), clouds tend to increase SNR more for both \ch{O2} and \ch{O3} when we assume small cloud particles. The cloud types that lead to the largest increase in SNR are stratus and deep convective, which both have high water content. Stratus clouds occur at low levels, and therefore enhance SNR for \ch{O2} and \ch{O3} similarly. Deep convective clouds extend throughout the troposphere, even to high levels, and therefore enhance \ch{O3} detection much more than \ch{O2}. Of the cloud types considered here, CARMA clouds are most similar to stratus \citep{yang2024impact}. Their effect on SNR is also similar to stratus, falling between the effect for stratus when we assume small and large particles. This is reasonable given that CARMA explicitly models cloud particle size and the resulting effective radii as a function of height generally fall between the small and large particle sizes assumed here \citep{yang2024impact}.

\section{Discussion}\label{sec:discussion}

First we compare our results with previous work on the impact of clouds on direct imaging detections by \citet{kawashima2019theoretical} and \citet{checlair2021theoretical}. Using Earth cloud reflectivity data, \cite{kawashima2019theoretical} found that for modern Earth levels of \ch{O2}, 100\% high cloud cover would increase the integration time needed for \ch{O2} detection by a factor of 10 (decrease SNR by a factor of 3) and 100\% low cloud cover would decrease the integration time needed for a detection an \ch{O2} by a factor of 10 (increase SNR by a factor of 3). In our sensitivity tests for \ch{O2} detection (Fig.~\ref{fig:sensi}b) with a typical cloud particle radius of 10~$\mu m$, a high cloud at 0.1~bar decreased the SNR by a factor of 2.0 and a low cloud at 1~bar increased the SNR by a factor of 2.2. This indicates that our methodology is roughly consistent with \citet{kawashima2019theoretical}, although the magnitudes of the cloud effects on SNR we report are somewhat smaller. \citet{checlair2021theoretical} simulated exoplanet atmospheres including an explicit representation of the vertical and horizontal distribution of clouds with ExoCAM \citep{wolf2022exocam}, a 3D GCM, then post-processed them using PSG to determine the SNR for \ch{O2} and \ch{O3} detection by reflected light spectroscopy. They found that clouds decrease the integration time to detect \ch{O2} by a factor of 2.0 and \ch{O3} by a factor of 2.3 with the \ch{O2} level of present Earth, which is equivalent to an increase by a factor of 1.4 for \ch{O2} and 1.5 for \ch{O3} in SNR. In this work, the experiments with baseline CARMA input shows that \ch{O3} SNR increases by a factor of 1.14 and \ch{O2} SNR increases by a factor of 2.0. This is the same scale as the 3-D results from \cite{checlair2021theoretical}, but reflects the impact of detailed cloud size distributions in this work and the impact of various cloud types in GCMs.

Compared to previous work \citep{kawashima2019theoretical, checlair2021theoretical}, our analytical model also provides a more thorough theoretical understanding of cloud impacts on biosignature detection. In previous work and in our PSG simulations, it was only possible to explore a limited set of biosignature gases and exoplanet parameters, but the analytical model allows us to generalize the conclusions to more situations. The limitations of this method should also be noted. Most importantly, our analytical model does not explicitly account for wavelength-dependent scattering and absorption effects, which can have a significant impact on detectability. For example, it does not include Rayleigh scattering by the atmosphere, which preferentially increases the albedo at short wavelengths and therefore enhances \ch{O3} detectability. The analytical model also does not include water vapor absorption, which occurs preferentially in the infrared, and therefore decreases \ch{O2} detectability. These factors are the main explanation for why \ch{O3} is much easier to detect than \ch{O2} in our PSG calculations (section~\ref{sec:planetary_params}). The specific vertical distribution of the background and target gas, for example, how concentrated \ch{O3} is distributed near the ozone layer, will also affect the conclusions quantitatively.

There are a number of aspects of our work that could be useful focuses of future inquiry. First, in our baseline experiments, we only included \ch{N2} and \ch{H2O} as background gases, but omitted low-concentration gases, such as \ch{SO2}, \ch{NO2}, \ch{NO}, \ch{CH4}, and \ch{CO2}. We performed tests to confirm that this assumption does not significantly affect our conclusions for modern-Earth concentrations of these gases, but they could have higher concentrations on some Earth-like exoplanets. Gases such as \ch{CH4} and \ch{CO2} that absorb mainly in the infrared are unlikely to affect \ch{O2} and \ch{O3} detection significantly even at high concentrations, but gases that have some overlap with important \ch{O2} and \ch{O3} absorption features might.

Second, we used the simple method of estimating SNR used by \citet{kawashima2019theoretical}, \citet{checlair2021theoretical}, \citet{checlair2021probing}, and \citet{afentakis2023understanding}, in which we compared the spectrum produced by all species to that produced by all species except the observational target (\ch{O2} or \ch{O3}). In a real retrieval an inversion with a fit to all relevant species at once would be necessary. This is particularly important for \ch{O2} and \ch{O3} because they have overlapping absorption bands and \ch{O3} is a photochemical product of \ch{O2}. Our work could be extended by including a photochemical model and treating the two gases as a combined biosignature.

Third, we assumed spherical cloud particles in the cloud microphysics model. This is a fair assumption for liquid droplets, but ice crystals can have a variety of shapes, especially hexagonal prisms. This can affect their optical properties and vertical transport dynamics. The ice particle aspect ratio and fractal dimension are a complex function of temperature and moisture. Future work could implement these aspects of cloud microphysics and determine whether they have any impact on the effect of clouds on \ch{O2} or \ch{O3} retrieval.

\section{Conclusion}\label{sec:summary}

In this work, we explored how clouds impact the reflected light direct imaging spectroscopic detection of biosignatures \ch{O2} and \ch{O3} on terrestrial planets. We used a variety of techniques to model relevant clouds, including running simulations with the 1D CARMA cloud microphysics model, in which we varied planetary parameters and imposed cloud data for different types of observed clouds on Earth. We fed these hypothetical cloud scenarios into PSG to simulate observations by HWO. We found that clouds are likely to increase the SNR for detections of \ch{O2} and \ch{O3} by HWO. Our work emphasizes that intuition gained from transmission spectroscopic observations may not translate simply to reflected light spectroscopy. 

\begin{acknowledgments}
Acknowledge:
This work was supported by NASA award No. 80NSSC21K1718, which is part of the Habitable Worlds program. This work was supported by NASA grant No. 80NSSC21K1533, which is part of the Future Investigators in NASA Earth and Space Science and Technology program.
\end{acknowledgments}


\end{document}